\documentclass[twocolumn,showpacs,preprintnumbers,amsmath,amssymb,aps]{revtex4}

\usepackage{graphicx}
\usepackage{dcolumn}
\usepackage{bm}

\usepackage{amsmath}
\usepackage{amscd}
\usepackage{amssymb}
\usepackage{bm}
\usepackage{eufrak}
\usepackage{mathrsfs}
\usepackage{lscape}
\usepackage{rotating}
\usepackage{caption2}
\usepackage{color}

\DeclareMathAlphabet{\mathsf}{T1}{cmss}{m}{sl}
\DeclareMathAlphabet{\mathsl}{T1}{cmss}{m}{sl}

\DeclareSymbolFont{lettersA}{U}{txmia}{m}{it}
\SetSymbolFont{lettersA}{bold}{U}{txmia}{bx}{it}
\DeclareFontSubstitution{U}{txexa}{m}{n}
 \makeatletter
   \DeclareMathSymbol{\m@thbbch@rC}{\mathord}{lettersA}{"83}
   \DeclareMathSymbol{\m@thbbch@rR}{\mathord}{lettersA}{"92}
 \makeatother

\newcommand{\ax}{\mathsl{a}}  
\newcommand{\A}{\mathsf{A}}
\newcommand{\Aa}{\ensuremath{\overset{\circ}
    {\mathbf{A}}}} 
\newcommand{\Aac}{\overset
   {\vbox{\moveright4pt\hbox{$\scriptstyle{\circ}$}}}{A}}
\newcommand{\Aap}[1]{\ensuremath{\overset{\circ}
     {\mathbf{A}}^{\lower3pt
      \hbox{$\scriptstyle{(\pi_{#1})}$}}}} %
\newcommand{\bara}{{\rlap/a}}
\newcommand{\bk}{{\rlap/\kappa}} 
\newcommand{\bkappa}{\bk}
\newcommand{\bl}{{\rlap/\lambda}} 
\newcommand{\blambda}{\bl}
\newcommand{\bxi}{{\bar\xi}}
\newcommand{\ex}{{\rm e}}
\newcommand{\F}{\digamma}  
\newcommand{\Fdual}
   {\mathsf{{}^*\kern-1.1pt \digamma}}  
\newcommand{\ga}{\mathsf{g}}
\newcommand{\m}{\hphantom{-}}
\newcommand{\miSigma}{\sigma}
\newcommand{\mik}[1]{\mathop{\overset{#1}{\bm\zeta}}\nolimits}
\newcommand{\mphi}{\mathit{\Phi}}  
\newcommand{\V}{V_{\kern-1pt 2}} 
\newcommand{\zb}{\bar z} 
\makeatletter
\newcommand{\Rreal}{
     \m@thbbch@rR
    }
\newcommand{\Ccompleja}{
     \m@thbbch@rC
    }
\makeatother

\newenvironment{eqn}[3][1]
{\begin{equation}\label{#3}
    \vcenter\bgroup \openup#1\jot
      \halign\bgroup #2\crcr}
{  \crcr\egroup\egroup
  \end{equation}%
}
\newenvironment{ecn}[2][$\displaystyle{####}$]
{\begin{eqn}{#1}{#2}}{\end{eqn}%
}

\begin{document}

\author{Tonatiuh Matos\email{tmatos@fis.cinvestav.mx}\footnote{Part
of the Instituto Avanzado de Cosmolog\'ia (IAC) collaboration
http://www.iac.edu.mx/}}
\email{tmatos@fis.cinvestav.mx}
\affiliation{Departamento de F\'{\i}sica,\\ Centro de Investigaci\'on y de Estudios Avanzados del
IPN, Apartado Postal 14-740, 07000 D.F, M\'exico .}
\author{Galaxia Miranda$^*$}
\email{galaxia@fis.cinvestav.mx}
\affiliation{Departamento de F\'{i}sica, \\ Escuela Superior de F\'{i}sica y
Matem\'{a}ticas del IPN,\\ Edificio 9, 07738 D.F., M\'{e}xico. } 
\author{Rub\'en S\'anchez-S\'anchez}
\email{rsanchez@ipn.mx}
\affiliation{Centro de Investigacion en Ciencia Aplicada y Tecnologia Avanzada del IPN,\\Legaria 694, 11500 D.F., M\'exico.}
\author{Petra Wiederhold}
\email{biene@ctrl.cinvestav.mx}
\affiliation{Departamento de Control Autom\'atico,\\ Centro de Investigaci\'on y de Estudios Avanzados del
IPN, Apartado Postal 14-740, 07000 D.F., M\'exico }

\title{Class of Einstein-Maxwell-Dilaton-Axion Space-Times}

\begin{abstract}
We use the harmonic maps ansatz to find exact solutions of
the Einstein-Maxwell-Dilaton-Axion (EMDA) equations. The solutions are harmonic maps invariant to the symplectic real group in four dimensions $Sp(4,\Rreal)\sim O(5)$.
We find solutions of the EMDA field equations for the one and two dimensional subspaces of the symplectic group. Specially, for illustration of the method, we find space-times that generalise the Schwarzschild solution with dilaton, axion and electromagnetic fields.
\end{abstract}

\draft


\date{\today}

\maketitle

\section{Introduction}\label{sec:Introduccion}

The new discoveries of the last years have changed our perspective and understanding of the Universe. Specially, the discovery of the dark matter and the dark energy have opened new big questions about the nature of the matter in Cosmos. Doubtless, it is time to propose new paradigms in order to give some light into these questions. One of the most accepted candidates to be the nature of the dark energy is a scalar field \cite{quintessence}, and maybe it is less known that scalar fields are also very good candidates to be the nature of the dark matter \cite{l9}. 

At the same time, theories like superstrings propose the existence of several scalar fields. In particular, at low energy the superstrings theory contains at least two scalar fields called the dilaton and the axion. There are some attempts looking for comparing these two scalar fields with the dark matter and dark energy \cite{linde} \cite{compean}, but the main problem for this is to go from the higher dimension theory to the four-dimensional one \cite{arbey}. In some cases it seems that this theory could explain the universe, but this question is still open.

In this work we study the Einstein-Maxwell-Dilaton-Axion (EMDA) system, from the effective point of view, i.e., we start from the corresponding Lagrangian and derive the field equations. Later we use the harmonic maps ansatz to solve the system of six coupled, non-linear differential equations for the axial symmetric stationary case.

The method of harmonic maps to find exact solutions of the
Einstein, Einstein-Maxwell and Einstein-Maxwell-Dilaton fields
has been used with great success. This ansatz was first
used by Neugebauer and Kramer to find exact solutions to
Einstein-Maxwell equations \cite{neu} and in \cite{ma24} this ansatz was generalised to the Einstein-Maxwell-Dilaton system with a coupling constant $\alpha$ between the dilaton and the Maxwell fields given by $\alpha=\sqrt{3}$. Later on this ansatz was generalised in \cite{maridari} for an arbitrary $\alpha$. The ansatz has been used also for solving the Einstein-Maxwell-Phantom system with arbitrary $\alpha$ \cite{ma87}. Here we apply the harmonic maps ansatz to solve
the equations of motion for the Einstein-Maxwell-Dilaton-Axion theory in the target space (see \cite{pollo}). 

This work is organised as follows. In section \ref{sec:La_accion} we introduce the fields of the potential space we are working with. In section \ref{sec:modelo_sigma} we write the field equations as a non-linear $\sigma$-model to be used in section \ref{sec:procedimiento}, where we use the harmonic maps ansatz to solve the system. In section \ref{sec:ejemplos1D} we solve the field equations for the one-dimensional subgroups of $Sp(4,\Rreal)\sim O(5)$ and in section \ref{sec:ejemplos2D} for the subgroup $SO(2,1)$. In section \ref{sec:conclusiones} some conclusions and perspectives are given. In the appendix \ref{App:HP} we review the use of the harmonic maps ansatz for the chiral equations, the non-linear $\sigma$-models.

\section{The effective action for EMDA}\label{sec:La_accion}
Gravity with two scalar fields, the dilaton and the axion and a $U(1)$ vector field can be described with the action
\begin{eqnarray}
    S&=&\frac{1}{16\pi}\int\left[ 
    -R+\frac{1}{3}\ex^{-4\mphi}H_{\mu\nu\lambda}H^{\mu\nu\lambda}\right.\nonumber \\
    &+&\left. 2\,\partial_\mu\mphi\,\partial^\mu\mphi-\ex^{-2\mphi}\F_{\mu\nu}
    \F^{\mu\nu} \right] \sqrt{-g}\,d^4x,
\end{eqnarray}
where we start with a space-time metric in four dimensions with the dilaton 
$\mphi$ coupled to the $U(1)$ vector field, the Maxwell field, with coupling constant $\alpha=1$ as in superstrings theory,
 such that $\F_{\mu\nu}=
\nabla_\mu \A_\nu - \nabla_\nu \A_\mu$ is the corresponding Maxwell Tensor plus the Pecci-Quinn pseudoscalar $\ax$. The Maxwell tensor can be written as $\F=d\A$. The antisymmetric tensor of three indices $H^{\mu\nu\lambda}$ is the Kalb-Ramon tensor defined as
\begin{eqnarray}
 H_{\mu\nu\lambda}&=&
       \Bigl(
       \partial_\mu B_{\nu\lambda} +
       \partial_\lambda B_{\mu\nu} + \partial_\nu B_{\lambda\nu}
       \Bigr)\nonumber\\
        &-&
       \Bigl(
       \A_\mu \F_{\nu\lambda} + \A_\lambda \F_{\mu\nu} +
       \A_\nu \F_{\lambda\mu}
       \Bigr).
\end{eqnarray}

In this description the electromagnetic 4-potential $\A_\mu$ has
two non zero components
 \[
       \A_\mu = \frac{1}{\sqrt{2}}(\psi,0,0,\sqrt{2}A_\varphi).
 \]
 On the other hand,
the Kalb-Ramon tensor has only one component $B_{03}=b$.

The symmetry group $Sp(4,\Rreal)$ for the EMDA model acts on the
set of the six potentials: $f$, the
gravitational; $\epsilon$, the rotational; $\psi$, the electrostatic; $\chi$, the magnetostatic; $\mphi$, the dilatonic and $\ax$ the axionic potential. The group $Sp(4,\Rreal)$ is homomorphic to the group $O(5)$, but in this work we will use the representations of $Sp(4,\Rreal)$. The three potentials $f,\psi$ and $\chi$ are dual to the three  potentials $\epsilon,\chi$ and $\ax$.
 Here $\ax$ is a Pecci-Quinn pseudoscalar field dual to the Kalb-Ramon tensor
$H^{\mu\nu\sigma}$
\[
      H^{\mu\nu\sigma} = \frac{1}{2} \ex^{4\mphi}
      E^{\mu\nu\sigma\tau}\frac{\partial\ax}
      {\partial x^\tau}.
\]
The effective action for the bosonic sector of a heterotic string
of ten dimensions compactified into four and with one vector
field $U(1)$ can be rewritten as
\begin{eqnarray}\label{la:accion}
     S&=&\frac{1}{16\pi}\int\Bigl\{
           -R+2\,\partial_\mu\mphi\,\partial^\mu\mphi
           +\frac{1}{2}\ex^{4\mphi}\partial_\mu\ax\,\partial^\mu\ax
            \nonumber\\
           &-&\ex^{-2\mphi}\F_{\mu\nu}\F^{\mu\nu}
           -\ax\,\F_{\mu\nu}\Fdual^{\mu\nu}
           \Bigr\}\,\sqrt{-g}\,d^4x.
\end{eqnarray}
Here $\Fdual= \frac{1}{2}E^{\mu\nu\lambda\tau}\F_{\lambda\tau}$ is
the dual of the Maxwell tensor. Also we have that
$E^{\mu\nu\lambda\tau} =
 \varepsilon^{\mu\nu\lambda\tau}{\mathrm{sign}}(g)/\sqrt{-g}$
is the Levi-Civita pseudo-tensor. To reduce the system to three
dimensions we need a non zero, time-like Killing vector. With this ansatz
it is possible to write the 4-dimensional metric $g_{\mu\nu}$ in
terms of the 3-dimensional $h_{ij}$ one as
\begin{eqnarray}\label{ansatz:4dim}
   ds^2 &=& ds_{(4)}{}^2
         =g_{\mu\nu}dx^\mu\,dx^\nu\nonumber\\
        &=&f(dt-\omega_i\,dx^i)^2-\frac{1}{f}h_{ij}dx^i\,dx^j.
\end{eqnarray}
(We use the convention:
Latin indices run in three dimensions, for example $i,j=1,2,3$ and
Greek indices run in four dimensions, for example
$\alpha,\beta=0,1,2,3$).
Here the three dimensional metric is given by
\begin{eqnarray}\label{tres:dim}
     ds_{(3)}{}^2&=&h_{ij} dx^i\,dx^j\nonumber\\
         &=& 2\,\ex^{2\Gamma}dz\,d\zb + \rho^2 d\varphi^2.
\end{eqnarray}
or, in Weyl coordinates we use the complex variable 
$z=\frac{1}{\sqrt{2}}(\rho + i\zeta) $, thus metric (\ref{tres:dim}) transforms into the Lewis-Papapetrou form
\[
  ds_{(3)}{}^2 = \ex^{2\Gamma}( d\rho^2 +d\zeta^2) + \rho^2 d\varphi^2.
\]
We will use also the Boyer-Lindquist coordinates $\rho =
\sqrt{r^2-2mr+\sigma^2}\sin(\theta)$ and $\zeta = (r-m)
\cos(\theta)$. In this coordinates the 3-metric (\ref{tres:dim}) reads
\begin{eqnarray}\label{ansatz:3dim}
     ds_{(3)}{}^2 &=& \ex^{2\Gamma}\biggl[
       \bigl((r-m)^2 +
       K^2\cos^2(\theta)\bigr)\Bigl(\frac{dr^2}{r^2-2mr+\sigma^2}
       +d\theta^2\Bigr)
       \biggr]\nonumber\\
&+& (r^2-2mr+\sigma^2)\sin^2(\theta) d\varphi^2.
\end{eqnarray}

The variation of the action \eqref{la:accion} gives the Euler-Lagrange equations for the fields, to obtain the following:
 the coupled Maxwell
equation with two scalar fields
\begin{equation}\label{Maxwell:mod}
   \nabla_\mu(\ex^{-2\Phi}\F^{\mu\nu}+\ax\,{}^*\F^{\mu\nu})=0,
\end{equation}
the dilaton and axion equations
\begin{gather}
  \nabla^\mu\nabla_\mu\mphi = \frac{1}{2}\ex^{-2\mphi}\F^2
    +\frac{1}{2}\ex^{4\mphi}(\partial\ax)^2,\label{ec:dilaton}\\
  \nabla_\mu(\ex^{4\mphi}g^{\mu\nu}\partial_\nu\ax)+
  \F_{\mu\nu}\,{}^*\F^{\mu\nu}=0,\label{ec:axion}
\end{gather}
and the main Einstein equations
\begin{eqnarray}\label{ec:Einstein}
    R_{\mu\nu} &=& 2\mphi_{,\mu}\mphi_{,\nu}+\frac{1}{2}\ex^{4\Phi}
    \ax_{,\mu}\ax_{,\nu}\nonumber\\
     &+& \ex^{-2\Phi}(2\F_{\mu\tau}\F^\tau{}_\nu
    + \frac{1}{2}\F^2g_{\mu\nu}).
\end{eqnarray}

If there exists a time-like Killing vector, it is possible to decompose the Maxwell tensor into two fields, the electrostatic $\psi$ and the magnetostatic $\chi$
potentials. With the help of these two quantities we can obtain
the electric $E_i=\F_{0i}$ and magnetic
$\F_{ij}=\varepsilon^{ijk}B_k$ components of the Maxwell tensor as
\begin{eqnarray}\label{pot:campos}
       \F_{i0}={}&\frac{1}{\sqrt{2}}\,\partial_i\psi, \label{eq:psi}\\
       \ex^{-2\Phi}\F^{ij}+\ax\,\Fdual^{ij}={}&\frac{f}{\sqrt{2h}}\,
       \varepsilon^{ijk}\partial_k\chi. \label{eq:chi}
\end{eqnarray}
The first relationship (\ref{eq:psi}) can
be deduced from the Bianchi identity
\[
      \nabla_\mu \Fdual^{\mu\nu} = 0.
\]

Another important quantity for this work is the twist 3-tensor $\tau_i$, this is derived from the rotational $\varepsilon$, the
magnetostatic $\chi$ and electrostatic $\psi$ potentials as
\[
      \tau_i = \partial_i\epsilon + \psi \partial_i\chi -
      \chi\partial_i\psi.
\]

The metric function $\omega_i=\omega_i(r,\theta)$ in the 4-metric in the Lewis-Papapetrou form \eqref{ansatz:4dim} is computed from
the relation
\[
      \tau^i =
      -\frac{f}{\sqrt{h}}\varepsilon^{ijk}\partial_j\omega_k.
\]
Thus, if we know the potentials, we can integrate the elements of the four-dimensional metric.

\section{The non-linear $\sigma$-model of the EMDA theory}\label{sec:modelo_sigma}

The most important feature we use here to find exact solutions for the EMDA field equations is the fact that the Euler-Lagrange equations \eqref{Maxwell:mod},
\eqref{ec:dilaton}, \eqref{ec:axion} and \eqref{ec:Einstein}, can
be obtained from the following action of the 3-dimensional non-linear
$\sigma$-model
\begin{eqnarray}
   S^{(3)} &=& \int\Bigl\{ R^{(3)} \nonumber\\
           &-&\frac{1}{2f^2}\bigl[(\nabla f)^2
     +(\nabla\epsilon+\psi\nabla\chi-\chi\nabla\psi)^2\bigr]
     -2(\nabla\mphi)^2\nonumber\\
     &-&\frac{1}{2}\exp{(4\mphi)}(\nabla\kappa)^2\nonumber\\
     &+&
      \frac{1}{f}\Bigl[\ex^{2\mphi}(\nabla\psi-\ax\nabla\chi)^2
     +\ex^{-2\mphi}(\nabla\psi)^2\Bigr]\Bigr\}\sqrt{h}
     d^3x.\nonumber\\
\end{eqnarray}
Alternatively this can be written as
\begin{equation}
    S_{(\sigma)} = \int ( R^{(3)} - \mathcal{G}_{AB}\,
    \partial_i \varphi^A \partial_j \varphi^B h^{ij} )
    \sqrt{h}\, d^3x,
\end{equation}
with the line element of the target space given by
\begin{eqnarray}\label{linea:potencial}
  dl^2 &=&\mathcal{G}_{AB}\,
    d \varphi^A d \varphi^B \nonumber \\
       &=&\frac{1}{2 f^2}\Bigl[df^2 + \bigl(
  d\epsilon + \psi d\chi - \chi d\psi \bigr)^2
  \Bigr]\nonumber\\ 
     &-& \frac{1}{f} \Bigl[ \ex^{2\mphi}
  \bigl(d\chi - \ax d\psi\bigr)^2
   + \ex^{-2\mphi} d\psi^2 \Bigr]\nonumber\\
  {} &+& 2\, d\mphi^2 + \ex^{4\mphi} d\ax^2,
\end{eqnarray}
where we have introduced the vector potential
\[
\varphi^A = \left( f,\epsilon,\psi,\chi,\mphi,\ax \right).
\]
This important line element can be derived from the following
Lagrangian density, which introduces the matrix $\ga\in
Sp(4,\Rreal)$ of potentials

\begin{equation}\label{lagrangiana:pot}
    \mathscr{L} = \frac{1}{4} \mathop{\rm Tr}\bigl(d\ga\,
    \ga^{-1}\,d\ga\,\ga^{-1}\bigr).
\end{equation}
in two dimensions. In terms of the complex variables $z$ and $\zb$
this is equivalent to
\[
    \mathscr{L} = \frac{1}{4}\mathrm{Tr}\biggl(
    \ga_{,z} \ga^{-1}_{,\zb} + \ga_{,\zb} \ga^{-1}_{,z} \biggr).
\]
The Euler-Lagrange equations of this relation are the chiral
equations
\begin{ecn}{quirales}
    (\ga_{,z} \ga^{-1})_{,\zb} + (\ga_{,\zb} \ga^{-1})_{,z} =0.
\end{ecn}

The form of $\ga$ can be expressed as a Gaussian decomposition of
$2\times2$ matrices $P$ and $Q$ given by
\begin{equation}\label{axion:dilaton:bloques}
  \ga=\left(\begin{array}{cc}
     P^{-1} & P^{-1}Q \\
     QP^{-1} & P + QP^{-1}Q
  \end{array}\right).
\end{equation}
where $P$ and $Q$ are
\begin{eqnarray}\label{P:Q}
   P&=&\left(\begin{array}{cc}
     f-\ex^{-2\mphi}\psi^2 & -\ex^{-2\mphi}\psi\\
     -\ex^{-2\mphi}\psi & -\ex^{-2\mphi}\hphantom\psi
   \end{array}\right),\quad\nonumber\\
   Q&=&\left(\begin{array}{cc}
      w\psi - \epsilon & \m w \\
      w & -\ax \end{array}\right),
\end{eqnarray}
here we have introduced the variable $w=\chi-\ax\psi$. Then solving the
quiral equation \eqref{quirales}, we can find solutions of the EMDA
theory.

\section{The Harmonic maps ansatz for $Sp(4,\Rreal)$-invariant chiral equations}\label{sec:procedimiento}

In this section we apply the harmonic maps ansatz explained in appendix \ref{App:HP} in order to solve the matrix equation \eqref{quirales}. Metric (\ref{linea:potencial}) defines a target space where the covariant derivatives of the Riemann tensor are zero. Thus, following the method given in appendix \ref{App:HP}, the Lie group element $\ga\in Sp(4,\Rreal)$ of the
topological Lie group $Sp(4,\Rreal)$ can be
parametrised in two variables $\xi$ and $\bxi$
as $\ga=\ga(\xi,\bxi)$. We know that since $Sp(4,\Rreal)$ is a
linear subgroup of $GL(n)$, then the Maurer-Cartan form
${\bm{\omega}}_{MC}$ on the tangent space $T_\ga(Sp(4,\Rreal))$ of $Sp(4,\Rreal)$, can
be defined by an element $\mathbf{v}_{\kern-2pt \ga}$ of
$T_\ga Sp(4,\Rreal)$ such that
\begin{equation}\label{Mau:Car}
      \Aa = \bm{\omega}_{MC}(\mathbf{v}_{\kern-2pt \ga}) =
      \mathbf{v}_{\kern-2pt \ga}\, \ga^{-1}.
\end{equation}
We can solve this equation to obtain 
  \begin{equation}\label{gauge:A}
       \ga_{,i}=\Aac_i(\ga)\,\ga,\qquad i=\xi,\bxi,
  \end{equation}
to get the matrix $\ga\in G$. It can be shown that if $\Aa$ is
built as
\begin{ecn}{la:A}
       \Aac_i(\ga) = \sum^{\mathrm{dim}\;\mathcal{G}_s}_{j=1}
       \overset{(k)}{\backepsilon_i}\,\varsigma_k,
\end{ecn}
being $\overset{(k)}{\backepsilon_i}$ Killing vectors of the
maximally symmetric space $\V$ with the two dimensional metric
\begin{equation}\label{auxiliar:esp}
      ds^2_{\V} = \frac{d\xi\,d\bxi}{V^2},
\end{equation}
where $V=1+k\xi\bxi$ and $\varsigma_k$ are the generators of the Lie algebra
$\mathcal{G}_s$ of the submanifold $G_s$ of $Sp(4,\Rreal)$. Then the element
$\ga\in Sp(4,\Rreal)$ of the exponential equation \eqref{gauge:A} is a
solution of the quiral equations \eqref{quirales} (see also Appendix \ref{App:HP}).

We find solutions of the EMDA problem, by solving the equations \eqref{gauge:A} in the two variables
$\xi$ and $\bxi$.

In our present case a representation of $\ga$ of the group $Sp(4,\Rreal)$ is given by \eqref{axion:dilaton:bloques} and \eqref{P:Q}.

\section{One-Dimensional Subspaces}\label{sec:ejemplos1D}

One-dimensional subspaces are the simplest subspaces to be handled and at the same time the richest ones. Therefore it is worth to study them with some deepness. In one dimension there is only one Killing-vector, thus the Killing equation (\ref{gauge:A}) reduces to solve the matrix equation
\begin{equation}
 \ga_{,\lambda}=A\,\ga,
\label{eq:g=Ag}
\end{equation}
where $\lambda$ is the parameter solution of the Laplace equation in one dimension
\begin{equation}
 (\rho\,\lambda_{,z})_{,\zb}+(\rho\,\lambda_{,\zb})_{,z}=0
\label{eq:Laplace}
\end{equation}
and $A\in sp(4,\Rreal)$, the corresponding Lie algebra of $Sp(4,\Rreal)$. Here it is convenient to use the fact that the chiral equations are invariant under the left action of the group $ Sp(4,\Rreal)$. Thus, if $B,D\in Sp(4,\Rreal)$ we have that $C=B\,A\,B^{-1}$ fulfils (\ref{eq:g=Ag}) with $\ga^{\prime}_{,\lambda}=C\,\ga^{\prime}$, being $\ga^{\prime}=B\ga D$. Then it is convenient to work with a representative of the equivalence class of $A$. It is easy to see that there are only two independent representatives of the equivalence class such that $A\in sp(4,\Rreal)$, the first one is
\begin{equation}
 A=\left( \begin{array}{llll}
           p & 0 & 0 & 0\\
0 & p^{\ast} & 0 & 0\\
0 & 0 & -p & 0\\
0 & 0 & 0 & -p^{\ast}
\end{array}\right) .\label{eq:A1}
\end{equation}
With matrix $A$ the solution of equation (\ref{eq:g=Ag}) is
\begin{equation}
 \ga=\left( \begin{array}{llll}
         A\,e^{p\lambda} & 0 & 0 & 0\\
0 & \frac{1}{B}\,e^{q\lambda} & 0 & 0\\
0 & 0 & \frac{1}{A}\,e^{-p\lambda} & 0\\
0 & 0 & 0 & B\,e^{-q\lambda}
\end{array}\right) 
\label{eq:ga1D}
\end{equation}
 We compare (\ref{axion:dilaton:bloques}) with (\ref{eq:ga1D}) to get the potentials
\begin{eqnarray}
 f&=&\frac{1}{A}e^{-p\lambda} \nonumber\\
  \nonumber\\
\mphi&=& \frac{1}{2}(q\lambda+\ln B) \nonumber\\
\psi&=& \epsilon=\chi=\ax=0
\label{eq:solI1}
\end{eqnarray}

Now we give some examples. If we take the solution of the Laplace equations (\ref{eq:Laplace}) $\lambda=\lambda_{0}\ln[1-\frac{2m}{r}]+m_{0}$, the potentials become
\begin{eqnarray}
 f&=&\frac{1}{Ae^{pm_{0}}}\left(
1-\frac{2m}{r}\right)  ^{-p\lambda_{0}}\nonumber\\
\phi&=&\frac{1}{2}\ln B+\frac{1}
{2}q\left(  m_{0}+\lambda_{0}\ln\left(1-2\frac{m}{r}\right)  \right)
\end{eqnarray}
The four dimensional space-time metric for this solution is then
\begin{eqnarray}
  ds^2 &=&  \frac{1}{f}\left[ {\hat K} dr^2 + (r^2-2mr)({\hat K}d\theta^2+\sin^2(\theta)\,d\varphi^2)\right] \nonumber\\
&-&f\,dt^2,
\end{eqnarray}
where
\[
 {\hat K}=\left( \frac{(r-m)^2 -
       m^2\cos^2(\theta)}{r^2-2mr}\right) ^{k_0}.
\]
For $r>>1$ this solution has the asymptotic behaviour given by 
\[
f\rightarrow
1+\frac{2mp\lambda_0}{r}+2m^{2}p(1+p)\frac{1}{r^{2}}+\cdots
\] 
and
\[
 \mphi\rightarrow\frac{1}{2}\ln B+\frac{1}{2}qm_{0}-\frac{q\lambda_0\,m}{r}+\cdots\hskip0.5cm .
\]

Another example is the following. We use now the harmonic map
\[
 \lambda=\lambda_{0}\ln\left( \frac{r-m-\sqrt{m^{2}-\sigma^{2}}}{r-m+\sqrt
{m^{2}-\sigma^{2}}}\right) +m_{0}.
\]
 In this case solution (\ref{eq:solI1}) becomes
\begin{eqnarray}
 f&=&\allowbreak\frac{1}{Ae^{pm_{0}}
}\left(  \frac{m-r+\sqrt{m^{2}-\sigma^{2}}}{m-r-\sqrt{m^{2}-\sigma^{2}}
}\right)  ^{-p\lambda_{0}}\nonumber\\
\mphi&=&\frac{q}{2}\left( \lambda_{0}\ln\left( \frac{r-m-\sqrt{m^{2}-\sigma^{2}}
}{r-m+\sqrt{m^{2}-\sigma^{2}}}\right) +m_{0}\right) +\frac{1}{2}\ln B\nonumber\\
\end{eqnarray}
and the four dimensional space-time metric for this solution is
\begin{eqnarray}
  ds^2 &=&  \frac{1}{f}\left[ {\hat K} dr^2 + (r^2-2mr+\sigma^2)({\hat K}d\theta^2+\sin^2(\theta)\,d\varphi^2)\right] \nonumber\\
&-&f\,dt^2,
\end{eqnarray}
with
\[
 {\hat K}=\left( \frac{(r-m)^2+(\sigma^2 -
       m^2)\cos^2(\theta)}{r^2-2mr+\sigma^2}\right) ^{k_0}
\]

Here the asymptotic behaviour for $r>>1$ is given by
\[
 f\rightarrow 1+\frac{2p\lambda_0\sqrt{m^{2}-\sigma^{2}}}{r}+\cdots
\]
and 
\[
 \mphi\rightarrow\frac{1}{2}\ln B+\frac{1}{2}qm_{0}-\frac{q\lambda_0\,\sqrt{m^2-\sigma^2}}{r}+\cdots,
\]
where we have set $Ae^{pm_{0}}=1$. We can use more harmonic maps in order to find more exact solutions. 

In what follows we study the second representative of $A\in sp(4,\Rreal)$,  given by
\begin{equation}
 A=\left( \begin{array}
[c]{cccc}%
p & 1 & 0 & 0\\
0 & p & 0 & 0\\
0 & 0 & -p & 0\\
0 & 0 & -1 & -p
\end{array}\right). \label{eq:A2}
\end{equation}
With this representative we obtain the solution
\begin{equation}
 \ga=\left( 
\begin{array}
[c]{cccc}
(a\lambda-a^{2}c)e^{p\lambda} & ae^{p\lambda} & 0 & 0\\
ae^{p\lambda} & 0 & 0 & 0\\
0 & 0 & 0 & \frac{1}{a}e^{-p\lambda}\\
0 & 0 & \frac{1}{a}e^{-p\lambda} & (-\frac{1}{a}\lambda+c)e^{-p\lambda}
\end{array}\right) 
\end{equation}
to obtain the potentials
\begin{eqnarray}
 f&=&\frac{e^{-p\lambda}}{a(\lambda-ac)}\nonumber\\
\phi&=&\frac{1}{2}\left[ p\lambda-\ln\left( \frac{1}{a}\lambda-c\right) \right] \nonumber\\
\psi&=&-\frac{1}{\lambda-ac}\nonumber\\
\epsilon&=&\chi=\ax=0
\label{eq:solI2}
\end{eqnarray}
This solution contains gravitational, dilaton and electrostatic fields, it represents a charged, dilatonic space-time. 
Nevertheless, in these two solutions (\ref{eq:solI1}) and (\ref{eq:solI2}), the axion field is zero. In order to find solutions with a non-zero axion field we perform the following procedure. Because the chiral equations are invariant under the left action of the group, we can perform a rotation $\ga^{\prime}\rightarrow C\ga C^{T}$, where $C^{T}$ means transpose of $C$. We start with the matrix 
\begin{equation}
 C=\left( 
\begin{array}
[c]{cccc}%
c & 0 & -b & 0\\
0 & c & 0 & -d\\
\frac{1}{b} & 0 & 0 & 0\\
0 & \frac{1}{d} & 0 & 0
\end{array}\right) 
\in Sp(4,R)
\end{equation}
and apply the left action of the group to the first representative \eqref{eq:A1}. If we do so, we obtain
\begin{widetext}
 \begin{equation}
 \ga^{\prime}=\left( 
\begin{array}
[c]{cccc}%
\frac{1}{A}e^{-p\lambda}\left(  b^{2}+A^{2}c^{2}e^{2p\lambda}\right)   & 0 &
\frac{A}{b}ce^{p\lambda} & 0\\
0 & \frac{1}{B}e^{q\lambda}\left(  c^{2}+B^{2}d^{2}e^{-2q\lambda}\right)   &
0 & \frac{1}{B}\frac{c}{d}e^{q\lambda}\\
\frac{A}{b}ce^{p\lambda} & 0 & \frac{A}{b^{2}}e^{p\lambda} & 0\\
0 & \frac{1}{B}\frac{c}{d}e^{q\lambda} & 0 & \frac{1}{Bd^{2}}e^{q\lambda}
\end{array}\right) 
\end{equation}
\end{widetext}

With matrix $\ga^{\prime}$ the physical potentials are
 \begin{eqnarray}
f&=& \frac{A\,e^{-p\lambda}}{A^{2}c^{2}+b^{2}e^{-2p\lambda}}\nonumber\\ \epsilon&=&-\frac{A^{2}\,c\,e^{2p\lambda}}{b^{3}+A^{2}bc^{2}e^{2p\lambda}
}\nonumber\\
e^{2\phi}&=&-\frac{1}{B}e^{q\lambda}\left(  c^{2}+B^{2}d^{2}e^{-2q\lambda
}\right)\nonumber\\
\ax&=&-\frac{c}{c^{2}d+B^{2}d^{3}e^{-2q\lambda}}\nonumber\\
w&=&\psi=\chi=0\nonumber\\
\label{eq:solI3}
\end{eqnarray}
Solution (\ref{eq:solI3}) represents a rotating, dilatonic solution coupled to an axion field. We show an example using the harmonic map $\lambda=\allowbreak m_{0}+\lambda_{0}\ln\left(  1-2\frac
{m}{r}\right) $. Substituting this $\lambda$ into the solution (\ref{eq:solI3}), we obtain
\begin{eqnarray}
f&=&\frac{A\,L_p^2}{ b^{2} +A^{2}c^{2}\,L_p^2} \nonumber\\
\epsilon&=&-\frac{A^{2}c\,L_p^2 }{b^{3}+A^{2}bc^{2}L_p^2}\nonumber\\
e^{2\phi}&=&-\frac{1}{B}c^{2}L_q-B\frac{d^{2}}{L_q}\nonumber\\
\ax&=&-\frac{c\,L_q^2}{B^{2}\,{d^{3}}+c^{2}d\,L_q^2}\nonumber\\
w&=&\psi=\chi=0\label{eq:SchwG1}
\end{eqnarray}
where 
\[
 L_p=e^{pm_{0}}  \left(  1-\frac{2m}{r}
\right)  ^{p\lambda_{0}}
\]
the four dimensional space-time metric for this solution is
\begin{eqnarray}
  ds^2 &=&  \frac{1}{f}\left[ {\hat K} dr^2 + (r^2-2mr+\sigma^2)({\hat K}d\theta^2+\sin^2(\theta)\,d\varphi^2)\right] \nonumber\\
&-&f\,(dt+a\,\cos(\theta)\,d\varphi)^2,
\end{eqnarray}
where
\[
 {\hat K}=\left( \frac{(r-m)^2 -
       m^2\cos^2(\theta)}{r^2-2mr}\right) ^{k_0}.
\]

The asymptotic behaviour for this solution $(r>>1)$ is given by

\begin{align}
f  &  \rightarrow1+\frac{4b^{2}mp\lambda_{0}e^{-2pm_{0}}}{Ar}+O(r^{-2}
)\nonumber\\
\epsilon &  \rightarrow-\frac{A\,c}{b}-\frac{4bcmp\lambda_{0}e^{-2pm_{0}}}
{r}+O(r^{-2})\nonumber\\
e^{2\phi}  &  \rightarrow-Bd^{2}e^{-\bar{p}m_{0}}-\frac{c^{2}e^{\bar{p}m_{0}}
}{B}-\frac{2re^{\bar
{p}m_{0}}m\bar{p}\lambda_{0}}{B}\nonumber\\
&+\frac{2Bd^{2}e^{-\bar{p}m_{0}}m\bar{p}\lambda_{0}}{r}+O(r^{-2})\\
{\ax}  &  \rightarrow-\frac{c\,e^{2\bar{p}m_{0}}}{B^{2}\,{d^{3}}
+c^{3}d\,e^{2\bar{p}m_{0}}}-\frac{4B^{2}cde^{2\bar{p}m_{0}}m\bar{p}\lambda
_{0}}{(B^{2}d^{2}+c^{3}e^{2\bar{p}m_{0}})^{2}r}+O(r^{-2})\nonumber
&
\end{align}
where $\frac{A\,e^{2pm_{0}}}{b^{2}+A^{2}c^{2}\,e^{2pm_{0}}}=1$. If we set $k_0=0$ and
\[
M=-\frac{2b^{2}mp\lambda_{0}e^{-2pm_{0}}}{A}
\] 
solution \eqref{eq:SchwG1} can be seen as a generalisation of the Schwarzschild space-time with rotation, dilaton and axion fields. Nevertheless, this solution is asymptotically flat only if $a=0$, when the solution becomes static.

In the same way we can apply the left action of the group to the second representative \eqref{eq:A2}. We use now the matrix
\begin{equation}
 C=\left( 
\begin{array}
[c]{cccc}%
0 & 1 & 0 & -1\\
-1 & 0 & 1 & 0\\
0 & 1 & 0 & 0\\
-1 & 0 & 0 & 0
\end{array}\right).
\in Sp(4,R)
\end{equation}
After the transformation $\ga^{\prime}\rightarrow C\ga C^{T}$, we obtain
\begin{widetext}
 \begin{equation}
 \ga^{\prime}=\left( 
\begin{array}
[c]{cccc}%
\frac{1}{A}\left(  ABe^{-p\lambda}-\lambda e^{-p\lambda}\right)   & \frac
{1}{A}e^{-p\lambda}\left(  -A^{2}e^{2p\lambda}-1\right)   & 0 & -Ae^{p\lambda
}\\
\frac{1}{A}e^{-p\lambda}\left(  -Aae^{2p\lambda}-1\right)   & A\lambda
e^{p\lambda}-A^{2}Be^{p\lambda} & -ae^{p\lambda} & A\lambda e^{p\lambda}
-A^{2}Be^{p\lambda}\\
0 & -Ae^{p\lambda} & 0 & -Ae^{p\lambda}\\
-ae^{p\lambda} & A\lambda e^{p\lambda}-A^{2}Be^{p\lambda} & -ae^{p\lambda} &
A\lambda e^{p\lambda}-A^{2}Be^{p\lambda}
\end{array}\right) 
\end{equation}
\end{widetext}

With this matrix we find the potentials
\begin{eqnarray}
 f&=& \frac{Ae^{p\lambda}}{AB-\lambda},\hskip0.5cm
\epsilon=0\nonumber\\
\psi&=&\frac{1-A^{2}e^{2p\lambda}}{AB-\lambda}\nonumber\\
\chi&=&-\frac{A^{2}e^{2p\lambda}}{AB-\lambda}\nonumber\\
e^{2\phi}&=&\frac{(A\,(2AB-\lambda)\lambda -A^{2}\,(2+AB^{2}))\,e^{2p\lambda}-A^{3}e^{4p\lambda}-A
}{(\lambda -AB)e^{p\lambda}}\nonumber\\
\ax&=&\frac{A^{2}\,((2AB-\lambda)\lambda-(1
+A^{2}B^{2}))-A^{4}\,e^{2p\lambda}}{(A^{4}+1)\,e^{2p\lambda}-A^{2}\,((2A\,B-\lambda)\lambda-2-A^{2}B^{2})}
\label{eq:solI4}
\end{eqnarray}

Metric (\ref{eq:solI4}) represents a dilatonic static space-time coupled to an axion field, with electric and magnetic charges. We can see explicitly this metric using some harmonic map $\lambda$. 
Again we only give an example with the harmonic map $\lambda=\allowbreak m_{0}+\lambda_{0}\ln\left(  1-\frac{2m}
{r}\right) $, using this in the solution we find
\begin{eqnarray}
 f&=& \frac{A\,L_p}{AB-m_{0}-\lambda_{0}\ln\left(  1-\frac{2m}{r}\right)  }\nonumber\\
\epsilon&=&0\nonumber\\
\psi&=&\frac{1-A^{2}L_p^2}{AB-m_{0}-\lambda_{0}\ln\left(
1-\frac{2m}{r}\right)  }\nonumber\\
\chi&=&-\frac{A^{2}\,L_p^2}{AB-m_{0}-\lambda_{0}\ln\left(  1-\frac{2m}{r}\right)  }\nonumber\\
e^{2\phi}&=&\frac{(A\,L_x -A^{2}\,(2+AB^{2}))\,L_p^2-A^{3}L_p^4-A
}{(m_{0}+\lambda_{0}\ln\left(  1-\frac{2m}{r}\right) -AB)\,L_p}\nonumber\\
\ax&=&\frac{A^{2}\,(L_x-(1
+A^{2}B^{2}))-A^{4}\,L_p^2}{(A^{4}+1)\,L_p^2-A^{2}\,(L_x-2-A^{2}B^{2})}
\end{eqnarray}
where
\begin{eqnarray}
 L_x&=&\left( 2AB-m_{0}-\lambda_{0}\ln\left(  1-\frac{2m}{r}\right)\right)\times\nonumber\\ &&\left(m_{0}+\lambda_{0}\ln\left(  1-\frac{2m}{r}\right)\right) 
\end{eqnarray}
The asymptotic behaviour for $r>>1$ for this solution is as follows. It is convenient to chose $A=\frac{m_0}{B-e^{pm_0}}$. In this case we have that
\[
 f\rightarrow 1-2\,{\frac {{\lambda_0}\,m \left( ({m_0}\,p-1){e^{p{
m_0}}}+B \right) }{{m_0}\,{e^{p{m_0}}}}}\frac{1}{r}.
\]
Again, if we define the mass parameter $M$ of this solution as
\[
 M=\,{\frac {{\lambda_0}\,m \left( ({m_0}\,p-1){e^{p{
m_0}}}+B \right) }{{m_0}\,{e^{p{m_0}}}}},
\]
the solution can be seen also as a generalisation of the Schwarzschild space-time. In this case, this solution has an electric monopole charge $Q$ 
\begin{eqnarray}
 Q&=&2\,\frac {{\lambda_0}\,m {e^{-p{m_0}}}}{{{m_0}}^{2} \left(B -{e
^{p{m_0}}} \right) }\left[ (1+2\,{{m_0}}^{3}p-{{m_0}}^{2}){e^{2\,p{m_0}}}\right. \nonumber\\
&-&\left. B \left((3\,-{{m_0}}^{2})\,{e^{p{m_0}}}+ {e^{-p{
m_0}}}{B}^{2}-3\,B
 \right) \right],
\end{eqnarray}
a dilatonic charge $Q_D$ given by
\begin{widetext}
 \begin{eqnarray}
 Q_D&=&\frac { m\,\lambda_0}{m_0\, \left( {B}^{2}+ \left( 2\,{m_0}\,{e^{2\,pm_0}}-2\,{e^{pm_0}} \right) B+2\,{m_0}^{2}
{e^{4\,pm_0}}-2\,m_0\,{e^{3\,pm_0}}+{e^{2\,pm_0}}
 \right) }\times\nonumber\\
&&\left[ -{e^{-p{m_0}}}{B}^{3}+. \left( 3+
 \left( p-2\,{e^{p{m_0}}} \right) {m_0} \right) {B}^{2}\right. \nonumber\\
 &+&\left. \left( 
-2\,{e^{2\,pm_0}}p{m_0}^{2}+ \left( 4\,{e^{2\,pm_0}}-2
\,p{e^{pm_0}} \right) m_0-3\,{e^{pm_0}} \right) B\right. \nonumber\\
&-&\left. 4\,{e
^{4\,pm_0}}p{m_0}^{3}+2\,{e^{3\,pm_0}}{m_0}^{2}p+
 \left( -2\,{e^{3\,pm_0}}+{e^{2\,pm_0}}p \right) m_0+{e^{2\,pm_0}} \right]
\end{eqnarray}
\end{widetext}
 and finally an axion $Q_{\ax}$ charge such that
\begin{widetext}
 \begin{eqnarray}
Q_{\ax}&=&\frac{4\, {m_0}^{2}\,m{\lambda_0}\,{e^{pm_0}} }{ \left( {e^{2\,pm_0}}{B}^{4}-4\,{e^{3\,
pm_0}}{B}^{3}+ \left( 6\,{e^{4\,pm_0}}+2\,{m_0}^{2}
 \right) {B}^{2}+ \left( -4\,{e^{5\,pm_0}}-4\,{m_0}^{2}{e^{p
m_0}} \right) B+2\,({m_0}^{2}+1)\,{e^{2\,p{
m_0}}}{m_0}^{2}+{e^{6\,pm_0}} \right) ^{2}}\times\nonumber\\
&&\left[   {B}^{6}{e^{pm_0}}p+ \left(m_0-6\,p \right)\, {e^{2\,pm_0}}{B}^{5}+ \left(15\,p-5\,m_0+{m_0}^{2}p \right)\,{e^{3\,pm_0}} {B}^{4} \right.  \nonumber\\
  &+&\left.  \left( (10\,m_0-20\,{p-4\,{m_0}^{2}p)\,e^{4\,pm_0}}+{m_0}^{3} \right) {B}^{3}\right.  \nonumber\\
  &+&\left.  \left( (6\,{m_0}^{2}p-10\,m_0+15\,p)\,{e^{5\,p{m_0}}}-({m_0}p+3)\,{m_0}^{3}\,{e^{pm_0}}\right) {B}^{2}\right.  \nonumber\\
  &+&\left.  \left( -(6\,p-4\,{m_0}^{2}p-5\,m_0)\,{e^{6\,pm_0}}+(3+2\,{{
m_0}}p)\,{m_0}^{3}\,{e^{2\,pm_0}} \right) B\right.  \nonumber\\
  &+&\left. {(e^{7\,p{
\it m0}}}{m_0}^{2}p+p-m_0)\,{e^{7\,pm_0}}-(1+{m_0}
p)\,{m_0}^{3}{e^{3\,pm_0}} \right] 
\end{eqnarray}
\end{widetext}

In order to see the physical behaviour of this solution, we take the very simple choice $m_0=0$, $B=1$. For this case, the asymptotic behaviour of the solutions for $r>>1$ is
\begin{eqnarray}
 f&=&  1-{\frac {2\,{\lambda_0}\,m\,(A\,p+1)}{A}}\frac{1}{r}+O \left({r}^{-2} \right) 
\nonumber\\
\psi&=&-{\frac { \left( A^2-1 \right) }{A}}+2\,{\frac {{
\lambda_0}\,m \left( 2\,{A}^{3}p+{A}^{2}-1 \right) }{{A}^{2}}}\frac{1}{r}+O \left({r}^{-2} \right) \nonumber
\end{eqnarray}
\begin{eqnarray}
\mphi&=&\frac{1}{2}\,\ln  \left( 1+2\,A+2\,{A}^{2} \right)\nonumber\\ &-&{\frac {{\lambda_0}\,m
 \left( 2\,{A}^{2}p+4\,{A}^{3}p+1+2\,A-pA \right) }{A \left( 1+2\,A+2
\,{A}^{2} \right) }}\frac{1}{r}+O \left({r}^{-2} \right) 
  \nonumber\\
\ax&=&  -{\frac {{A}^{2} \left( 1+2\,{A}^{2} \right) }{2\,{A}^{4}+1+2\,{A}^{2}}}\\
&+&4\,{\frac {{\lambda_0}\,m{A}^{2} \left( -A-{A}^{3}-{A}^{2}p+p{A}^
{4}-p \right) }{ \left( 1+2\,{A}^{2}+2\,{A}^{4} \right) ^{2}}}
\frac{1}{r}+O \left({r}^{-2} \right)\nonumber
\end{eqnarray}
This behaviour shows the quantities we can consider as to be the physical parameters of the solution. This metric is an asymptotically flat generalisation of the Schwarzschild space-time, it is static and contains electromagnetic, scalar and axion parameters.

We can choose now other harmonic maps to generate more solutions, but the important point is that we can generate solutions with physical features we want to have. The same situation happens with the two dimensional subspaces that we will study in the next section.

\section{Two-Dimensional Subspaces: the subgroup $SO(2,1)$}\label{sec:ejemplos2D}
For this subgroup we start with the base of the algebra
$so(2,1)$ given by
\begin{eqnarray}\label{base:so21}
   \vcenter{\openup5\jot
   \halign{
          &\hfil$#$\cr
     \miSigma
     _1&=&\left(\begin{array}{cccc}
       \m0&\m0&\m1&\m0\hspace{4pt}\\
       \m0&\m0&\m0&\m1\hspace{4pt}\\
        -1&\m0&\m0&\m0\hspace{4pt}\\
       \m0& -1&\m0&\m0\hspace{4pt}\\
       \end{array}\right),
        &\qquad\nonumber\\
     \miSigma_2&=&\left(\begin{array}{rrrr}
       \m0&\m0&\m0&-1\hspace{4pt}\\
       \m0&\m0&-1&\m0\hspace{4pt}\\
       \m0&-1&\m0&\m0\hspace{4pt}\\
       -1&\m0&\m0&\m0\hspace{4pt}\\
       \end{array}\right),\nonumber\\
     \miSigma_3&=&\left(\begin{array}{cccc}
       \m0&\m1&\m0&\m0\hspace{4pt}\\
       \m1&\m0&\m0&\m0\hspace{4pt}\\
       \m0&\m0&\m0& -1\hspace{4pt}\\
       \m0&\m0& -1&\m0\hspace{4pt}\\
       \end{array}\right).\cr
  }
 }
\end{eqnarray}
We take the following Killing basis of the maximally symmetric
space $\V$:
\begin{eqnarray}\label{eq:kill123}
    \mik1&=&\frac{C_1}{V^2}
   (\blambda k\bar\xi^2+\bar\blambda,\,
    \bar\blambda k\xi^2+\blambda); \cr
    \mik2&=&\frac{i\>C_2}{V^2}\rlap/a(-
    \bar\xi,\xi);\hfill\cr
    \mik3&=&\frac{C_3}{V^2}(\bar\blambda k
    \bar\xi^2+\blambda,\,\blambda k\xi^2+\bar\blambda),
\end{eqnarray}
where $\blambda=\bkappa+i\bkappa\in\Ccompleja;
 \quad\bkappa\in\Rreal$. Now we choose the 
Maurer-Cartan form as
\begin{ecn}{des1:A}
    \Aap3=\mik{2}\miSigma_1+\mik{1}\miSigma_2+\mik{3}\miSigma_3.
\end{ecn}
The integrability condition $\ga_{,\xi\bxi} =\ga_{,\bxi\xi}$ for
$\ga\in Sp(4,\Rreal)$ is fulfilled provided that the
constants $C_j$, $j=1,2,3$ are restricted to
\begin{eqnarray}\label{so21n213Cs1}
   C_1&=&-\frac{i}{\bkappa}\sqrt{\frac{k}{2}},\nonumber\\
   C_2&=&\m\frac{2k}{\bara},\nonumber\\
   C_3&=&-\frac{i}{\bkappa}\sqrt{\frac{k}{2}}.
\end{eqnarray}
Thus after solving \eqref{gauge:A}, we find a solution of the
potential matrix $\ga$, given in \eqref{axion:dilaton:bloques} and
\eqref{P:Q}. Remember the fact that $\ga$ is a real and symmetric matrix, thus the conditions
for symmetry $\ga=\ga^t$ and reality $\ga\in Sp(4,\Rreal)$ must be taken into account. With this in main we get a solution for the potential matrix $\ga$, to obtain
\begin{equation}
 \ga=\left(\begin{array}{llll}
            \Xi & \Pi & 0 & \Pi \\
            \Pi & \Xi & \Pi & 0 \\
            0 & \Pi & \Xi & \Pi \\
            \Pi & 0 & \Pi & \Xi \\
\end{array}\right) 
\end{equation}
where \[
       \Xi=\frac{1-k\xi\bxi}{1+k\xi\bxi}
      \]
and \[
     \Pi=\displaystyle \sqrt{\frac{k}{2}}\frac{\m(1-i)\xi-(1+i)\bxi}{1+k\xi\bxi}.
    \]
With this solution we find the following set of potentials:
\begin{eqnarray}\label{potenciales:1}
   f &=& \frac{1-\xi\bxi}{1+\xi\bxi},\nonumber\\
   \epsilon &=& 0,\nonumber\\
   \psi &=&
   \frac{i}{\sqrt{2}}\frac{((i-1)\xi+(1+i)\bxi)}{1+\xi\bxi},\nonumber\\
   \chi &=& -\frac{1}{W}(\xi+\bxi+i\xi\bxi^2+i\bxi-i\xi+\xi^2\bxi+
   \xi\bxi^2-i\bxi\xi^2-i\xi^3)\nonumber\\
    &-&\frac{1}{W}(\bxi^3\xi^2+i\bxi^3+\xi^3\bxi^2-i\xi^3\bxi^2-
   \xi^3-\bxi^3+i\xi^2\bxi^3),\nonumber\\
   \ax &=& -\frac{\xi^2+\bxi^2}{1+\xi^2\bxi^2-i\xi^2+i\bxi^2},\nonumber\\
   \mphi &=&
   -\frac{1}{2}\ln\left(\frac{1-\xi^2\bxi^2}{-1-\xi^2\bxi^2+i\xi^2-i\bxi^2}
   \right).
\end{eqnarray}
where $W={\sqrt{2}}\,(-i\xi^3\bxi+i\bxi^2-i\xi^2+\xi^2\bxi^2+
   i\xi\bxi^3+\xi^3\bxi^3+\xi\bxi+1)$. 
The functions $\xi$ and $\bxi$ are solutions of the two dimensional harmonic maps equations, that means, of the Ernst equations \cite{kra}. We show an example using the Ernst potential for the Kerr solution,
with the help of an harmonic map defined in terms of the Ernst's potential
\begin{equation}\label{ec:map:1}
   {\mathcal{E}}=\frac{1-\xi}{1+\xi},
\end{equation}
by
\begin{equation}\label{ec:map:2}
   {\mathcal{E}}= 1+\frac{2q}{r-il\cos(\theta)}.
\end{equation}

We find the solution of the EMDA field equations such that
\begin{align}
    & ds^2 =
    f^{-1}\Biggl(dt
   -\frac{2q^2l\sin(\theta)^2}{r^2+2qr+l^2\cos(\theta)^2}\,d\varphi\Biggr)^2
    \notag \\ &
   -f\,
    \Biggl\{\:
      \biggl(1+\frac{q^2\sin(\theta)^2}
      {r^2+2qr+l^2\cos(\theta)^2}\biggr)^{-2} \label{sol1:repe}
     \\ & \qquad
      \bigl((r+q)^2+(l^2-q^2)\cos(\theta)^2\bigr)\,
      \biggl(\frac{dr^2}{\Delta}+d\theta^2\biggr)\\ &
      + \Delta\sin(\theta)^2\,d\varphi^2
    \:\Biggr\}, \notag
\end{align}
where $\Delta$ designs the horizon function, which is defined
by 
\[
     \Delta(r)\equiv r^2+2qr+l^2,
\]
and $f$ is the gravitational potential which reads
\[
 f=\biggl(1+\frac{2q^2}{r^2+2qr+l^2\cos(\theta)^2}\biggr).
\]

The harmonic map which is responsible of this solution is given by
\eqref{ec:map:1} and \eqref{ec:map:2}, or by
\begin{equation}\label{mapeo:armonico}
    \xi=\frac{m}{\mathsf{R}},\qquad\quad
\end{equation}
where  $\mathsf{R}\equiv r+q-il\cos(\theta)
    \in \Ccompleja$. 
This harmonic map satisfies the harmonic equations, which in complex coordinates read
\begin{ecn}{miarmonica}
   (\rho\xi_{,z})_{,\zb} + (\rho\xi_{,\zb})_{,z} +
   2\rho\Gamma^{\xi}{}_{\xi\xi}\xi_{,z}\xi_{,\zb} = 0,
\end{ecn}
where $\Gamma^{\xi}{}_{\xi\xi}$ are the affin connection of the
auxiliar space \eqref{auxiliar:esp}.

\begin{figure}
\begin{center}
\includegraphics[width=2.5in]{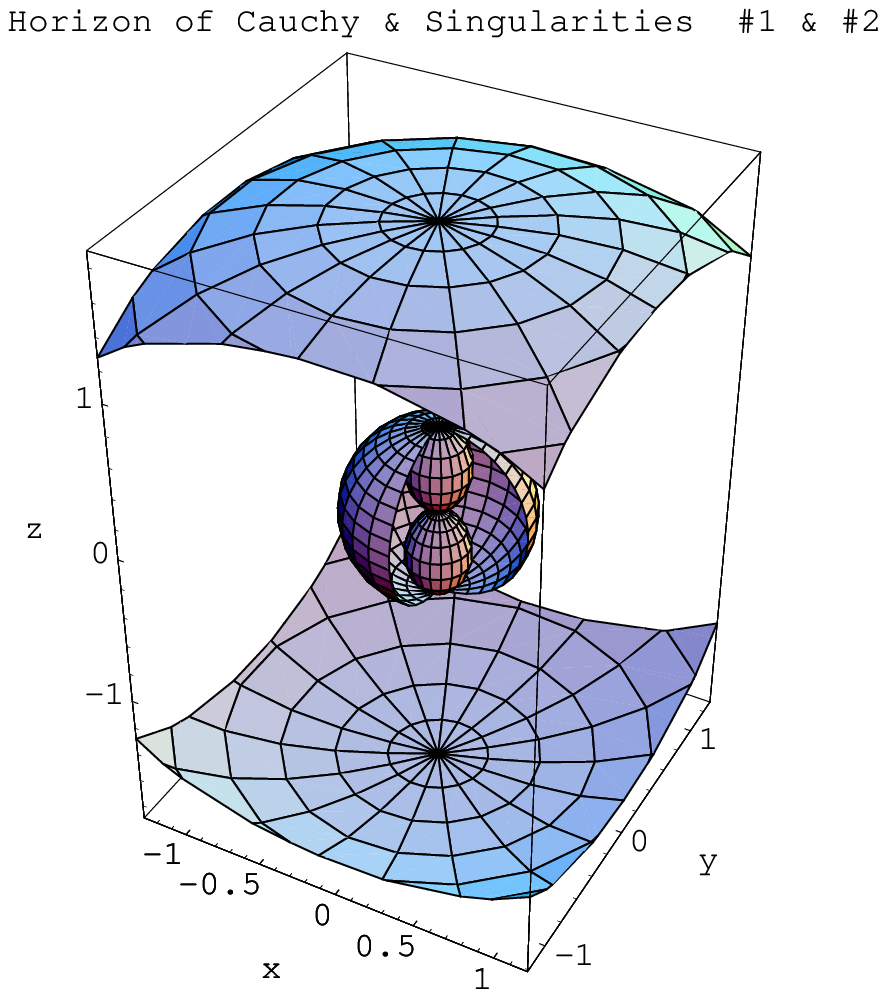}
\vskip20pt
\includegraphics[width=2.5in]{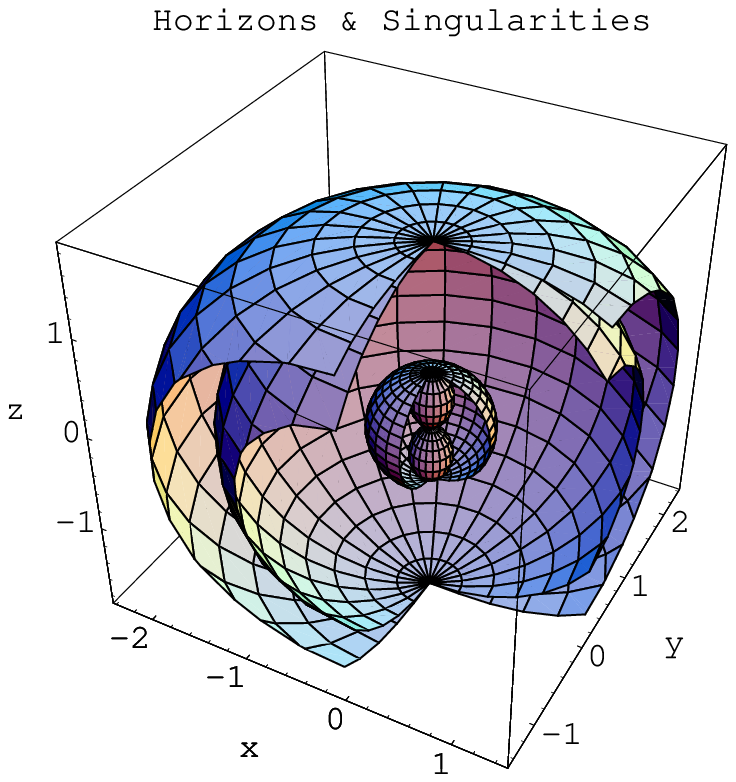}
\end{center}
\caption[Horizons and singularities]{Diagrams showing several
views of the horizons (spherical surfaces) $r=m\pm\sqrt{m^2-a^2}$
and the singularities (external surface like an ellipsoid $\#1$,
and internal surface $\#2$) $r=m\pm\sqrt{m^2-a^2\cos(\theta)^2}$,
for the values $m=1.18$ and $a=1$ in spherical coordinates.}
\label{fig:gr4vara}
\end{figure}

The other interesting aspect of the solution is the
electromagnetic field, this is 
\begin{align}
   E_r ={}&{}
   -\frac{q\bigl(r^2+2qr-2l\cos(\theta)(r+q)-l^2\cos^2(\theta)\bigr)}
   {\bigl(r^2+2qr+2q^2+l^2\cos^2(\theta)\bigr)^2},\notag\\
   E_\theta ={}&
   \m\frac{ql\sin(\theta)\Upsilon_{+}}
   {\bigl(r^2+2qr+2q^2+l^2\cos^2(\theta)\bigr)^2},\notag\\
   E_\varphi ={}& \m0,\notag\\
   B_r ={}& \m\frac{(r^2+2qr+l^2+2q^2)\, q\sin(\theta)\Upsilon_{-}}{
   \bigl(r^2+2qr+2q^2+l^2\cos^2(\theta)\bigr)^2}\mathop�
   {},\notag\\
   B_\theta ={}&
   \m\frac{\bigl((r+2q)r-2a(r+q)\cos(\theta)-l^2\cos^2(\theta))
    ql\sin^2(\theta)}{\bigl(r^2+2qr+2q^2+l^2\cos^2(\theta)\bigr)^2},
    \notag \\
   B_\varphi ={}& \m0.\label{EB1}
\end{align}
being 
\[
 \Upsilon_{\pm}=r^2+2qr+2q^2
   -l^2\cos^2(\theta)\pm2l(r+q)\cos(\theta).
\]

Using the Komar integrals we can find
the electric and the magnetic charges of the solution. If the electric charge is $q$ we find that the magnetic monopolar
charge is $p=-q$, which tell us that the solution is a dyon. The
parameter $l$, which is responsible of the stationarity of the metric is 
the parameter of dipolar electric moment.

Another feature of this solution is that it is asymptotically flat. The Komar mass is null $M=0$ and its angular moment $J=0$ too. Of course, this analysis is valid only from the point of
view of an observer that is far away from the source of the fields.

The solution contains two singularities with two regions separated in two geometric places

\begin{enumerate}
 \item the exterior singularity
  \[
 r = q + \sqrt{ q^2 - l^2\cos(\theta)^2 },
\]
 \item the interior singularity
\[
  r = q - \sqrt{ q^2 - l^2\cos(\theta)^2},
\]
\end{enumerate}

with horizons in

\begin{enumerate}
 \item the exterior horizon (Events)
\[
 r_{+} = q + \sqrt{ q^2 - l^2 },
\]
\item the interior horizon (Cauchy)
\[
 r_{-} = q - \sqrt{ q^2 - l^2 }.
\]
\end{enumerate}

The surface gravity of the exterior horizon is given by
\[
   \kappa = \frac{1}{\sqrt{2}}\frac{m^2-a^2}{m^2a},
\]
which tell us that it is a regular events horizon.

The solution is then a dyon, which represents a collapse of
electromagnetic charges. 
That latter fact follows from the nature of the coupling between gravity and the two scalar fields: the dilaton and the Pecci-Quinn pseudoscalar or
axion.

\section{Conclusions}\label{sec:conclusiones}

The harmonic maps ansatz is an excellent
tool for finding exact solutions of systems of non-linear partial
differential equations \cite{ma24}, in particular, this method has been very useful in solving the chiral equations derived from a non-linear $\sigma$ model \cite{ma29}. Einstein equations in vacuum can be reduced to a non-linear $\sigma $ model with structural group $SL(2,R)$ in the space-time and to a structural group $SU(1,1)$ in the potential spaces, $i.e.$, in terms of the Ernst potentials. The electro-vacuum case can also be reduced to a non-linear $\sigma $ model with structural group $SU(2,1)$ in terms of the extended Ernst potentials \cite{neu}, \cite{kra}. The Kaluza-Klein field equations can also be written as a $SL(3,R)$ non-linear $\sigma $ model in the space-time as well as in the potential space \cite{ma1}, \cite{ma24}. This is possible because the corresponding potential space is a symmetric Riemannian space only for $\alpha =0$ and $\alpha =\sqrt{3}$, but this is not the case for the low energy limit in superstrings or the Maxwell-phantom theories. In \cite{maridari} we extended this method \cite{ma24}, \cite{mis} to the Einstein-Maxwell-Dilaton fields with arbitrary $\alpha$ and in this work we use this technique for the Einstein-Maxwell-dilaton-axion fields with the invariant group $Sp(4,\Rreal)$. With this method we were able to obtain exact solutions of the EMDA field equations for the one- and two-dimensional subgroups of $Sp(4,\Rreal)$. The method is very powerful, it makes possible to generalise the Schwarzschild space-time and to obtain solutions which represent magnetic and electric monopoles, dipoles, dyons, etc., coupled to gravitational monopoles, dipoles and to different multipoles of the scalar fields.

\section*{Acknowledgements}\label{sec:agradecimientos}
We would like to thank
Dar\'{\i}o N\'u\~nez for
many helpful and useful discussions. The numerical computations were
carried out in the "Laboratorio de Super-C\'omputo
Astrof\'{\i}sico (LaSumA)" del Cinvestav and in the UNAM's cluster Kan-Balam. This work was partially
supported by CONACyT M\'exico, under grants 49865-F and I0101/131/07 C-234/07, Instituto Avanzado de Cosmologia (IAC) collaboration.

\appendix

\section[]{The Harmonic Maps Ansatz}\label{App:HP}

In this appendix we follow \cite{ma29} in order to apply the method of harmonic maps to the EMDA field equations.

Let $g$ be a map defined by
\begin{eqnarray*}
 g: {\mathcal C} \otimes {\bar{\mathcal C}} &\rightarrow& G \\ 
g &\rightarrow& g (z, \bar z) \in G,
\end{eqnarray*}
where $G$ is a paracompact Lie group and $g$ fulfils the field equations derived from the Lagrangian
\begin{equation}
 {\mathcal L} = \alpha\, \text{tr}(g_{,z} g^{-1} g_{,\bar z} g^{-1}). 
\label{eq:2}
\end{equation}
The field equations are called the chiral equations for $g$, in explicit form they are given by
\begin{equation}
 (\alpha g_{,z} g^{-1})_{,\bar z} + (\alpha g_{,\bar z}
g^{-1})_{,z} =  0 \label{eq:1}
\end{equation}
where $\alpha^2 = \det g$.

Lagrangian (\ref{eq:2}) represents a topological quantum field theory with gauge group $G$. In what follows we give a method to find explicit expressions of the
elements $g \in G$ in terms of the local coordinates $z$ and $\bar
z$.

Let $G_c$ be a subgroup $G_c \subset G$ such that $c\in G_c$ implies
$c_{,z} = 0$, $c_{,\bar z} = 0$. Then equation (\ref{eq:1}) is invariant under the left action $L_c$ of
$G_c$ over $G$.

Proposition 1. Let $\beta$ be a complex function defined by
\begin{equation}
 \beta_{,z} = {1\over 4(\ln\alpha)_{,z}}\, \text{tr}(g_{,z} g^{-1})^2, \ \ \
g\in G 
\end{equation}\label{eq:3}
and $\beta_{,\bar z}$ with $\bar z$ insted of $z$. If $g$ fulfils
the chiral equations, then $\beta$ is integrable.

Proof. It is sufficient to calculate the identity $\beta_{,z\bar z} = \beta_{, \bar zz}.$ To show this, we see that

\begin{eqnarray*}
 \beta_{,z\bar z}&=&\frac{1}{4}\, \text{tr}\left[ \frac{1}{\alpha_{,z}}  (\alpha g_{,z} g^{-1})_{,\bar z}\, g_{,z} g^{-1}+\frac{1}{\alpha_{,z}} \alpha g_{,z} g^{-1}\, g_{,z\bar z} g^{-1}\right. \\
&-&\left.\frac{1}{\alpha_{,z}}\alpha g_{,z} g^{-1} g_{,z} g^{-1}g_{,\bar z}g^{-1}- \frac{\alpha_{,z\bar z}}{(\alpha_{,z})^2}\alpha g_{,z} g^{-1} g_{,z} g^{-1}\right]\\
&=&\frac{1}{4}\frac{1}{\alpha_{,z}}\, \text{tr}\left[  \left(  (\alpha g_{,z} g^{-1})_{,\bar z}+(\alpha g_{,\bar z} g^{-1})_{,z}\right.\right.  \\
&-&\left.\left. \alpha_{, z}  g_{,\bar z}g^{-1}\right) g_{,z} g^{-1} \right]  
\end{eqnarray*}
but the matrices in the trace can be commutated. Thus, if $g$ fulfils the chiral equations, we have
\begin{equation*}
  \beta_{,z\bar z}=-\frac{1}{4}\, \text{tr}\left[    g_{,\bar z}g^{-1} g_{,z} g^{-1} \right].
\end{equation*}\hfill$\blacksquare$

Let ${\mathcal G}$ be the corresponding Lie algebra of $G$. The Maurer-Cartan
form $\omega_g$ of $G$ defined by 
\[
 \omega_g = L_{g^{-1}*} (g)
\]
is a one-form
on $G$ with values on ${\mathcal G}, \omega_g \in T^*_g G \otimes
{\mathcal G}$, where $T_x^* M$ represents the tangent space of the manifold $M$ at the point $x$ and $L$ is the left action of $G$ over $G$, $L: G \otimes G \rightarrow G$. $L$ must be defined in a convenient manner in order to preserve the properties of the elements of $G$. Now we define the mappings

\begin{eqnarray}
 A_z:G &\rightharpoonup &{\mathcal G}\nonumber\\ 
g &\rightharpoonup & A_z (g)  = 
g_{,z} g^{-1}\nonumber\\
 A_{\bar z}:G &\rightharpoonup& {\mathcal G}\nonumber\\ 
g &\rightharpoonup &
A_{\bar z} (g)  =  g_{,\bar z} g^{-1}
\label{eq:4}
\end{eqnarray}

If $g$ is given in a representation of $G$, then we can write the one-form
$\omega (g) = \omega_g$ as

\begin{equation}
 \omega = A_z dz + A_{\bar z} d {\bar z}. \label{eq:5}
\end{equation}

We can now define a metric on ${\mathcal G}$ in a standard manner. Since
$\omega_g$ can be written as in (\ref{eq:5}), the tensor

\begin{equation}
 l = \text{tr} (dgg^{-1} \otimes dgg^{-1}) \label{eq:6}
\end{equation}
on $G$ defines a metric on the tangent bundle of $G$.

Theorem 1. The submanifold of solutions of the chiral equations 
$S\subset G$, is a symmetric manifold with metric (\ref{eq:6}).

Proof. We will
only outline here the proof. We take a parametrisation $\lambda^a \,\, a=1, \cdots n$ of $G$. The
set $\{\lambda^a\}$ is a local coordinate system of the $n$-dimensional
differential manifold $G$. In terms of this parametrisation the Maurer-Cartan
one-form $\omega$ can be written as
\begin{equation}
 \omega = A_a d\lambda^a, \label{eq:7}
\end{equation}
where $A_a(g) = (\frac{\partial}{\partial \lambda^a} g) g^{-1}.$ The chiral
equations then read
\begin{equation}
 \nabla_b A_a (g) + \nabla_a A_b (g) = 0, \label{eq:8}
\end{equation}
with $\nabla_a$ the convariant derivative defined by (\ref{eq:6}).

It follows the relation
\begin{equation}
 \nabla_b A_a (g) = {1\over 2} \ [A_a, A_b] (g). \label{eq:9}
\end{equation}

Thus the Riemannian curvature ${\mathcal R}$ can be derived from (\ref{eq:6}), their components read
\begin{equation}
 R_{abcd} = \frac{1}{4} \text{tr} (A_{[a}A_{b]} A_{[c}A_{d]}) \label{eq:10}
\end{equation}
where $[a,b]$ means index commutation. This can be done, because $G$ is a
paracompact manifold. From here it follows that $\nabla {\mathcal R}= 0$.\hfill$\blacksquare$

Proposition 2. The function $\alpha^2 =\det g$ is harmonic.

Proof. Using the formulae tr$(A_{,x}A^{-1})=\ln(\det A)_{,x}$ we can see that the trace of the chiral equations implies $\alpha_{,z\bar z} =
 0.$  \hfill$\blacksquare$

Let $V_p$ be a complete totally geodesic submanifold of $G$ and let $\{
\lambda^i \} \, \, \, i = 1, \cdots , p$ be a set of local coordinates on $V_p$ and suppose we completely know the submanifold $V_p$. It is clear that the
submanifold $V_p$ is also symmetric. The symmetries of $G$ and $V_p$ are
in fact isometries, since both of them are paracompact manifolds, with Riemannian metrics (\ref{eq:6}) and $i_*l$ respectively, where $i$ is the restriction of $V_p$ into $G$. Let us suppose that $V_p$ possesses $d$ isometries. The chiral equations imply
\begin{eqnarray}
 (\alpha \lambda^i_{,z})_{,\bar z} + (\alpha \lambda^i_{,\bar
z})_{,z} + 2 \alpha \sum _{ijk} \Gamma^i_{jk} \lambda^j_{,z} \lambda^k_
{,\bar z} =  0, \nonumber\\
 i, j, k = 1, \cdots  ,p\label{eq:11}
\end{eqnarray}
where $\Gamma^i_{jk}$ are the Christoffel symboles of $i_*l$ and $\lambda^i$
are the totally geodesic parameters on $V_p$. In terms of the parameters
$\lambda^i$ the chiral equations read
\begin{equation}
  \nabla_i A_j (g) + \nabla_j A_i (g) = 0 \label{eq:12}
\end{equation}
where $\nabla_i$ is the covariant derivative of $V_p$. Equation (\ref{eq:12}) is the
Killing equation on $V_p$ for the components of $A_i$. Since we know the
manifold $V_p$, we know its isometries and therefore its Killing vector
space. Let $\xi_s,\,\,  s=1,\cdots , d, $ be a base of the Killing vector space
of $V_p$ and $\Gamma^s$ be a base of the subalgebra corresponding to $V_p$.
Then we can write
\begin{equation}
 A_i (g) = \sum _s\, \xi^i_s \, \Gamma^s \label{eq:13}
\end{equation}
where $\xi_s  = \sum _j \, \xi^j_s \, {\partial \over \partial \lambda^j}.$ The
covariant derivative on $V_p$ is given by
\begin{equation}
 \nabla_j A_i (g) = - {1\over 2} [A_{i,} A_j] (g) \label{eq:14}
\end{equation}
where $A_i$ fulfills the integrability conditions
\begin{equation}
 F_{ij} = \nabla_j A_i (g) - \nabla_i A_j (g) - [A_{j,}A_i](g) = 0
\label{eq:15}
\end{equation}
$i.e.$, $A_i$ has a pure gauge form.

Thus, because we know $\{\xi_s\}$ and $\{\Gamma^s\}$ we can integrate the elements of
$S$, since $A_i (g) \in {\mathcal G}$ can be mapped into the group by means
of the exponential map. Nevertheless it is not possible to map all the
elements one by one. Fortunately we have the following proposition.

Proposition 3. The relation $A^c_i \sim A_i$ iff there exist $c\in
G_c$ such that $A^c = A \circ L_c$, is an equivalence relation.

This equivalence relation separates the set $\{A_i\}$ into equivalence
classes $[A_i]$. Let $TB$ be a set of representatives of each class,
$TB = \{[A_i]\}$. Now we map the elements of $TB \subset {\mathcal G}$ into the
group $S$ by means of the exponential map or by integration. Let us define
$B$ as the set of elements of the group, mapped from each representative
\[
 B = \{g \in S|g = \exp (A_i), A_i \in TB\}\subset G.
\]
The elements
of $B$ are also elements of $S$ because $A_i$ fulfils the chiral equations,
$i.e.$ $B \subset S$. For constructing all the set $S$ we have the following theorem.

Theorem 2. $(S, B, \pi , G_c, L)$ is a principal fibre bundle with
projection $\pi(L_c(g)) = g$; $(c,g) = L_c (g)$.

Proof. The fibres of $G$ are the orbits of the group $G_c$ on $G,F_g =
\{g^\prime \in G|g^\prime = L_c (g)\}$ for some $g\in B$. The
topology of $B$ is its relative topology with respect to $G$. Let ${\cal B}_F$
be the bundle ${\cal B}_F = (G_c \times U_{\alpha , } U_{\alpha}, \pi)$, where
$\{U_\alpha\}$ is an open covering of $B$. We have the following lemma.

Lemma 1. The bundle ${\cal B}_F$ and 
\[
 {\cal B} = (\pi^{-1} (U_\alpha),
U_{\alpha ,}\pi|_{\pi^{-1}(U_a)})
\]
are isomorphic.

Proof. The mapping
\begin{equation*}
 \psi_\alpha: \phi^{-1} (U_\alpha) = \{g\in S|g^\prime = L_c (g),
g \in U_\alpha\}_{c\in G_c} \rightarrow G_c \times U_\alpha
\end{equation*}
\[
 g^\prime \rightarrow \psi_\alpha (g^\prime) = (c,g)
\]
is a homeomorphism and 
\[
 \pi|_{\pi^{-1}(U_{\alpha})} (g^\prime) = g = \pi_2
\circ \psi_\alpha (g^\prime).
\]
\hfill$\blacksquare$

By lemma 1 the bundle $\cal B$ is locally trivial. To end the proof of the
Theorem it is sufficient to prove that the $G_c$ spaces $(S,G_c, L)$ and
$(G_c \times U_{\alpha ,} G_{c,}\delta)$, are isomorphic, but that follows from
\[
 \delta \circ \text{id}|_{G_c} \times \psi_\alpha = \psi_\alpha\circ L|_{G_c
\times \pi^{-1}(U_\alpha)}.
\]
\hfill$\blacksquare$

With this theorem it is now possible to explain the harmonic maps method as follows:
\begin{itemize}
 \item Given the chiral equations (1), invariant under the group $G$, choose a
symmetric Riemannian space $V_p$ with $d$ Killing vectors, $p \leq n =$ dim
$G$.
\item Look for a representation for the corresponding Lie Algebra ${\mathcal G}$
compatible with the commutating relations of the Killing vectors, via
equation (14).
\item Write the matrices $A_i(g)$ explicitly in terms of the geodesic parameters
of the symmetric space $V_p$.
\item Use proposition 2 for finding the equivalence classes in $\{A_i\}$ and
choose a set of representatives.
\item Map the lie algebra representatives into the group.
\end{itemize}
The solutions can be constructed by means of the left action of the $G_c$
group into $G$.


\end{document}